\documentclass[superscriptaddress,longbibliography,aps,prl,reprint,showpacs]{revtex4-1}

\usepackage[english]{babel}
\usepackage[T1]{fontenc}
\usepackage{graphicx}
\usepackage{amsmath}
\usepackage{soul}
\usepackage{natbib}
\usepackage{eqnarray}
\usepackage{textcomp}
\usepackage{xcolor}

\bibstyle{aapmrev4-1}
\begin{document}
\title{Finite size effects on the ultrafast remagnetization dynamics of FePt}

\author{L. Willig}
\affiliation{Institut f\"ur Physik \& Astronomie,  Universit\"at Potsdam, Karl-Liebknecht-Str. 24-25, 14476 Potsdam,  Germany}
\affiliation{Helmholtz Zentrum Berlin, Albert-Einstein-Str. 15, 12489 Berlin, Germany}

\author{A. von Reppert}
\affiliation{Institut f\"ur Physik \& Astronomie,  Universit\"at Potsdam, Karl-Liebknecht-Str. 24-25, 14476 Potsdam,  Germany}

\author{M. Deb}
\affiliation{Institut f\"ur Physik \& Astronomie,  Universit\"at Potsdam, Karl-Liebknecht-Str. 24-25, 14476 Potsdam,  Germany}

\author{F. Ganss}
\affiliation{Institut f\"ur Physik, Technische Universit\"at Chemnitz, Reichenhainer Str. 70, 09126 Chemnitz, Germany}

\author{O. Hellwig}
\affiliation{Institut f\"ur Physik, Technische Universit\"at Chemnitz, Reichenhainer Str. 70, 09126 Chemnitz, Germany}
\affiliation{Institut f\"ur Ionenstrahlphysik und Materialforschung, Helmholtz-Zentrum Dresden-Rossendorf, Bautzner Landstrasse 400, 01328 Dresden, Germany}
	
\author{M. Bargheer}
\email{bargheer@uni-potsdam.de}
\homepage{http://www.udkm.physik.uni-potsdam.de} \affiliation{Institut  f\"ur Physik \& Astronomie, Universit\"at Potsdam,  Karl-Liebknecht-Str. 24-25, 14476 Potsdam, Germany}
\affiliation{Helmholtz Zentrum Berlin, Albert-Einstein-Str. 15, 12489 Berlin, Germany}

\newcommand{\superscript}[1]{\ensuremath{^{\textrm{#1}}}}
\newcommand{\subscript}[1]{\ensuremath{_{\textrm{#1}}}}

\date{\today}
\begin{abstract}
We investigate the ultrafast magnetization dynamics of FePt in the L$1_0$ phase after an optical heating pulse, as used in heat assisted magnetic recording. We compare continuous and nano-granular thin films and emphasize the impact of the finite size on the remagnetization dynamics. The remagnetization  speeds up significantly with increasing external magnetic field only for the continuous film, where domain wall motion governs the dynamics. The ultrafast remagnetization dynamics in the continuous film are only dominated by heat transport in the regime of high magnetic fields, whereas the timescale required for cooling is prevalent in the granular film for all magnetic field strengths. These findings highlight the necessary conditions for studying the intrinsic heat transport properties in magnetic materials.
\end{abstract}

\maketitle
The fascinating field of ultrafast magnetization dynamics has developed rapidly from the first demonstration of femtosecond demagnetization\,\cite{Beaurepaire1996} towards all-optical switching\,\cite{Stanciu2007, Lambert2014}, magnetization reversal by ultrashort electron pulses\,\cite{Yang2017} and heat-assisted magnetic recording (HAMR)\,\cite{Challener2009,Weller2016, Casoli2016_7}. Future light-based data-recording applications motivate an understanding of the fundamental light-matter-interactions in magnetic materials with nanoscale bit-dimensions, which are a key to satisfy the increasing demand for high-density information storage technology.

Information storage devices using the HAMR scheme have demonstrated that data densities beyond 1.4Tb/in$^{2}$ are feasible since long-term data stability of nanoscopic bits can be achieved by using materials with large perpendicular anisotropy\,\cite{Hono2018, Weller2013}. FePt in the highly ordered L1$_\text{0}$ phase is a promising material since it combines a large uniaxial magneto-crystalline anisotropy $K_\text{u}\cong7\cdot$10$^{6}$\,J/m$^{3}$, with a relatively low Curie temperature ($T_\text{C}\leq 750$\,K) and the possibility to grow nanograins with diameters down to 3\,nm by commercially viable sputtering techniques\,\cite{Weller2013}.

Besides nanoscopic bit volumes for high information density, it is of utmost importance to write and read information at the fastest possible speed and with the highest efficiency. This has triggered many research projects studying the ultrafast magnetization response of thin films to rapid heating via pulses of optical photons, photoinduced hot electrons or few picosecond long electrical currents.\,\cite{Beaurepaire1996, Gorchon2017,Mendil2014,Zhao2010}. Much of the fundamental research has focused on the timescales and mechanisms of the spin-angular momentum transfer during the demagnetization process\,\cite{Koopmans2009,Bigot2009,Battiato2010}. The subsequent field-assisted remagnetization has received less experimental attention despite its equal importance for future working devices\,\cite{Kazantseva2008}.

Magnetization switching in continuous thin films under thermal equilibrium requires the nucleation, propagation and finally coalescence of reversed magnetic domains. The substantial laser heating used in the HAMR-scheme adds the need for understanding nanoscale thermal transport and potential changes of material properties such as the magnetic anisotropy that occur upon heating close to and above $T_\text{C}$, as in fascinating all-optical switching experiments, which report toggle switching of the magnetization in the absence of an external field\,\cite{Gorchon2017}.

Current modeling approaches for the remagnetization dynamics\,\cite{Mendil2014, Chimata2012, Nieves2016, Lyberatos2019} often employ a three temperature model for the equilibration of electron-, phonon- and spin-temperatures where the local spin-temperature defines a stochastic magnetic field, that enters in an atomistic Landau-Lifshitz-Gilbert (LLG) equation. Extending the atomistic LLG-model to micromagnetic simulations of the macrospin evolution using the Landau Lifshitz-Bloch equation computes the magnetization dynamics on the relevant picosecond to nanosecond timescale even for macroscopic specimens\,\cite{Mendil2014, Nieves2016}. The incorporation of temperature dependent material parameters\,\cite{Lyberatos2019, Scheid2019}, quantized spin-states\,\cite{Nieves2016} and the proper quantum-thermodynamics noise-distribution function\,\cite{Barker2019} further improve the comparison to experiments. However, spatial confinement effects and interfaces that alter the domain propagation and the thermal transport are often not implemented in the modeling.

Here we present an experimental comparison of the magnetization dynamics of equally thin continuous and nano-granular FePt after medium to high-fluence femtosecond laser-excitation. The observed remagnetization-process speeds up by a factor of three with increasing external field for the continuous film. This effect is significantly reduced for the nano-granular FePt and for medium fluences the timescale of remagnetization is independent of the external field. We argue that domain wall propagation is irrelevant in the grains which are magnetically decoupled by the surrounding amorphous carbon-matrix. In the continuous film the remagnetization rate saturates for high external fields of about 0.7\,T and it approaches the timescale observed in the granular film, where the remagnetization speed is limited by the heat dissipation rate. We discuss the influence of the carbon matrix, which rapidly absorbs about 30$\%$ of the deposited energy and partially stores it for about 1\,ns.

\begin{figure}
	\centering
	\includegraphics[width = 8.6cm]{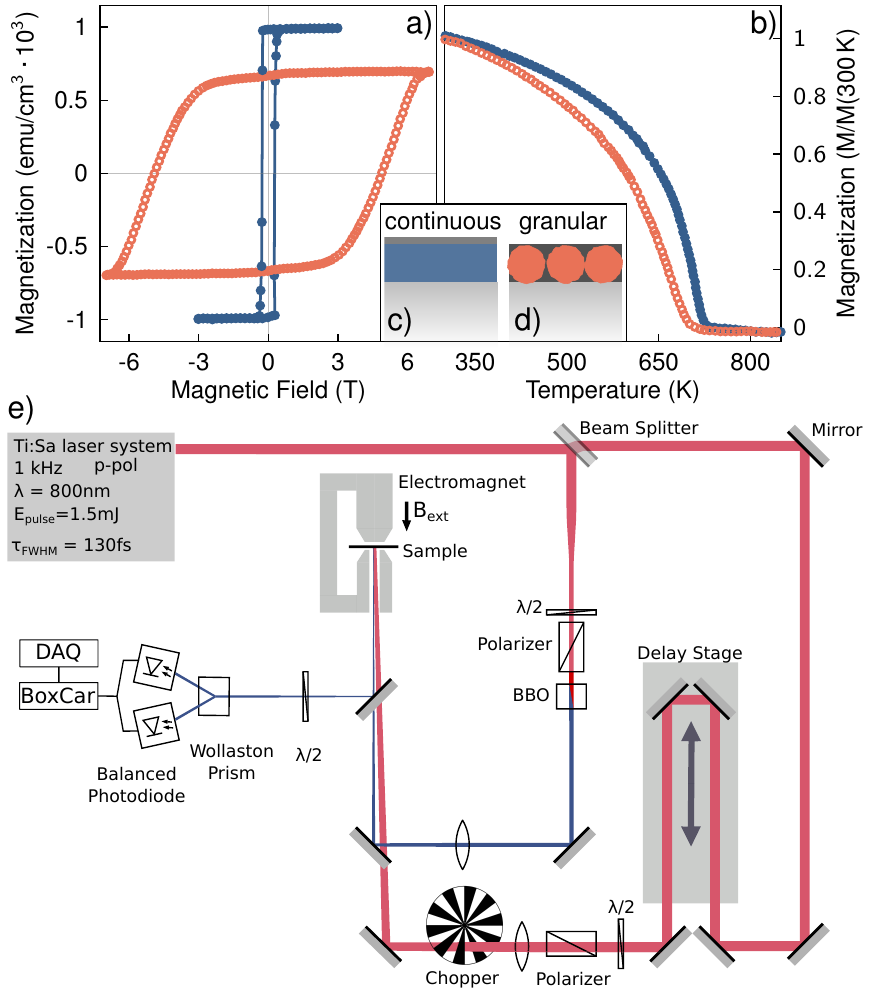}
	\caption{Static magnetic properties of the FePt samples and sketch of the TR-MOKE setup. (a),(b) Magnetization hysteresis loops (a) and temperature dependence of the magnetization (b) of the granular (open symbols) and continuous (plain symbols) samples. The magnetic field was applied perpendicular to the plane of the samples.(c),(d) Sketch of the continuous (c) and the granular (d) FePt samples. (e) Sketch of the TR-MOKE setup.}
	\label{fig:static}
\end{figure}

The investigated continuous and granular FePt films are grown in the ordered L1$_\text{0}$ phase onto MgO (100) oriented substrates, which aligns the easy magnetic axis out of plane. A more detailed description of the growth conditions and the structural properties of the samples can be found in Ref\,\cite{Reppert2018}. In particular, we mention that the granular film consists of segregated FePt-nanograins embedded in amorphous carbon with a size distribution of FePt particles centered at approximately 10\,nm (see Fig.~\ref{fig:static}\,(d)). The continuous film of FePt is capped by a 1\,nm Al layer, which is oxidized to Al$_2$O$_3$ (see Fig.~\ref{fig:static}\,(c)). The static magnetic properties of the samples were characterized using a superconducting quantum interference device - vibrating sample magnetometer under an external magnetic field ($B_\text{ext}$) applied normal to the sample plane. The measured hysteresis loops of the continuous and granular films are shown in Fig.~\ref{fig:static}\,(a). Their square shape shows that the magnetic easy axis is oriented perpendicular to the plane of the films. The coercive field of the granular film is around 5\,T, which is very large compared to the coercive field of the continuous film of approximately 0.4\,T. This is due to the different magnetization reversal mechanisms. Domain wall nucleation and propagation governs the continuous film, whereas the magnetization dynamics in the nanoparticles are dominated by quasi-coherent magnetization reversal \cite{Casoli2016_5}. In addition, we observed a reduction of the saturation magnetization ($M_\text{S}$) by 30\% in the granular film, which can be related to the volume occupied by the nonmagnetic carbon matrix in that film together with a small contribution related to the well-known effect of finite size\cite{Lyberatos2013} on $M_\text{S}$. The temperature dependence of the saturation magnetization was also measured in a wide range of temperatures between 300 and 850\,K (see Fig.~\ref{fig:static}\,(b)) for similarly prepared samples. The Curie temperature of the granular film is about $\cong$660\,K, approximately 30\,K lower than the value obtained for the continuous film, which is also in agreement with previous investigations of the finite size effect on $T_\text{C}$\,\cite{Hovorka2014}.
The laser induced ultrafast magnetization dynamics was investigated using the time-resolved  magneto-optical Kerr effect (TR-MOKE) setup sketched in Fig.~\ref{fig:static}\,(e). The pump and probe pulses were generated from an amplified Ti:Sapphire laser system delivering 130\,fs pulses centered at 800\,nm at the repetition rate of 1\,kHz. The pump beam is kept at the fundamental of the amplifier at 800\,nm and excites the sample under a small angle from the surface normal to the sample ($\cong$2\textdegree), while the probe beam is frequency doubled to 400\,nm with a nonlinear Beta-Barium-Borate      (BBO) crystal and incident onto the sample at almost perpendicular incidence ($\cong$1\textdegree). A chopper reduces the repetition rate of the pump pulses to 500\,Hz enabling pulse-to-pulse comparison of the pumped and unexcited states of the sample at the full 1\,kHz repetition rate. Both light pulses are linearly polarized and focused through one of the pole-shoes of an electromagnet onto the sample on a spot of 1000\,$\mu$m for the pump and 300\,$\mu$m for the probe. The reflected probe pulses allow measuring the differential change of the polar Kerr rotation ($\Delta\theta_{K}$) using a polarization bridge consisting of a $\lambda$/2 waveplate, a Wollaston prism and a balanced photodiode. The detected signal is analyzed via a Boxcar integrator and a data acquisition (DAQ) card. The $B_\text{ext}$ of the electromagnet is applied perpendicular to the surface of the sample.

\subsection{ Results and discussion}
\begin{figure}
	\includegraphics[width = 8.6cm]{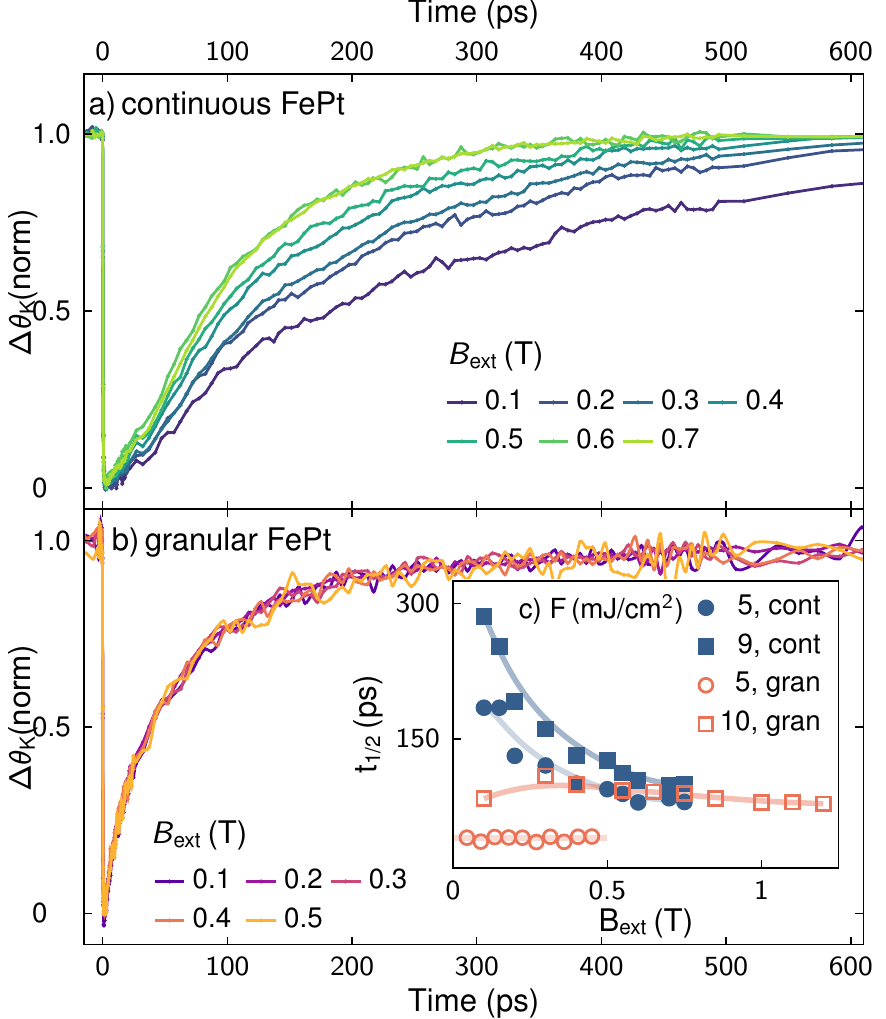}
	\caption{Magnetic field dependence of laser induced magnetization dynamics. (a,b) Normalized Kerr rotation $\Delta\theta_{K}$ measured in the (a) continuous and (b) granular samples at different applied magnetic fields for a fixed pump fluence of 5\,mJ/cm$^2$. The inset (c) shows the magnetic field dependence of the time $t_\text{1/2}$ at which half of the demagnetization amplitude has recovered for this pump fluence of 5\,mJ/cm$^2$ and an approximately doubled fluence of 9 and 10\,mJ/cm$^2$ for the continuous and granular sample, respectively. Solid lines are a guide to the eye.}
	\label{fig:field}
\end{figure}

Figure 2 shows the TR-MOKE measured for the continuous (Fig.~\ref{fig:field}\,(a)) and the granular (Fig.~\ref{fig:field}\,(b)) samples as a function of $B_\text{ext}$ at a pump fluence of $F_\text{pump}$=5\,mJ/cm$^{2}$. To compare the field effect on the ultrafast magnetization dynamics, $\Delta\theta_{K}$(t) signals were scaled by the $\Delta\theta_\text{K}$ amplitude corresponding to the maximum demagnetization. In both samples, the pump laser pulses induce a subpicoseond demagnetization process, which is independent of $B_\text{ext}$. We focus on the remagnetization process following this ultrafast demagnetization. Interestingly, a clear difference in the effect of $B_\text{ext}$ on this remagnetization is observed for the continuous and granular films. Indeed, the remagnetization of the continuous film speeds up significantly when the external field increases from 0.1 to 0.7\,T (see Fig~\ref{fig:field}\,(a)), while the  remagnetization of the granular one is independent of $B_\text{ext}$ (see Fig.~\ref{fig:field}\,(b)). In order to study this phenomenon in more detail, we measured the TR-MOKE at high pump fluence of $F_\text{pump}$=10\,mJ/cm$^{2}$ and over a wide range of $B_\text{ext}$ up to 1.2\,T. The results of this study are summarized in Fig.~\ref{fig:field}(c). For both samples and at the two $F_\text{pump}$ values of 5\,mJ/cm$^{2}$ and 10\,mJ/cm$^{2}$ we show the field dependence of the time $t_\text{1/2}$ corresponding to the time in which half of the demagnetization amplitude has recovered. Interestingly, the effect of $B_\text{ext}$ on the remagnetization becomes more pronounced in the continuous film at higher fluence and it reaches a saturation around 0.6 and 0.7\,T, while the remagnetization of the granular film remains weakly sensitive to $B_\text{ext}$, even though the range of applied fields is nearly twice as large.
It is straightforward to assign the large field dependence of the continuous film to domain-wall propagation that is well-known to govern the magnetization dynamics in continuous thin films and very sensitive to the amplitude of $B_\text{ext}$\,\cite{Cowburn1999, Metaxas2007, Hubert2011}. On the other hand, the domain wall propagation is irrelevant in nanosized magnets where the remagnetization should be governed by the cooling of the grain. Furthermore, we show that $t_\text{1/2}$ in the continuous film converges under high $B_\text{ext}$ to a value of $\cong$100\,ps, similar to the one characterizing the granular film at high fluence. Such saturation can be explained by the fact that the remagnetization in the continuous film at high fields is essentially governed by the dissipation of heat which cools the FePt spin system. Indeed, when the external field is large enough, it can keep the entire film essentially in a monodomain state such that domain wall propagation does not play a role. Fig.~\ref{fig:field}(a) clearly shows that only at high magnetic fields the heat transport is the dominant process for the ultrafast remagnetization dynamics in the continuous magnetic film. Therefore, in order to study the intrinsic heat transport properties in magnetic materials, it is necessary to investigate the magnetization dynamics in nanosized structures or under a high external magnetic field.

\begin{figure}
	\includegraphics[width = 8.6cm]{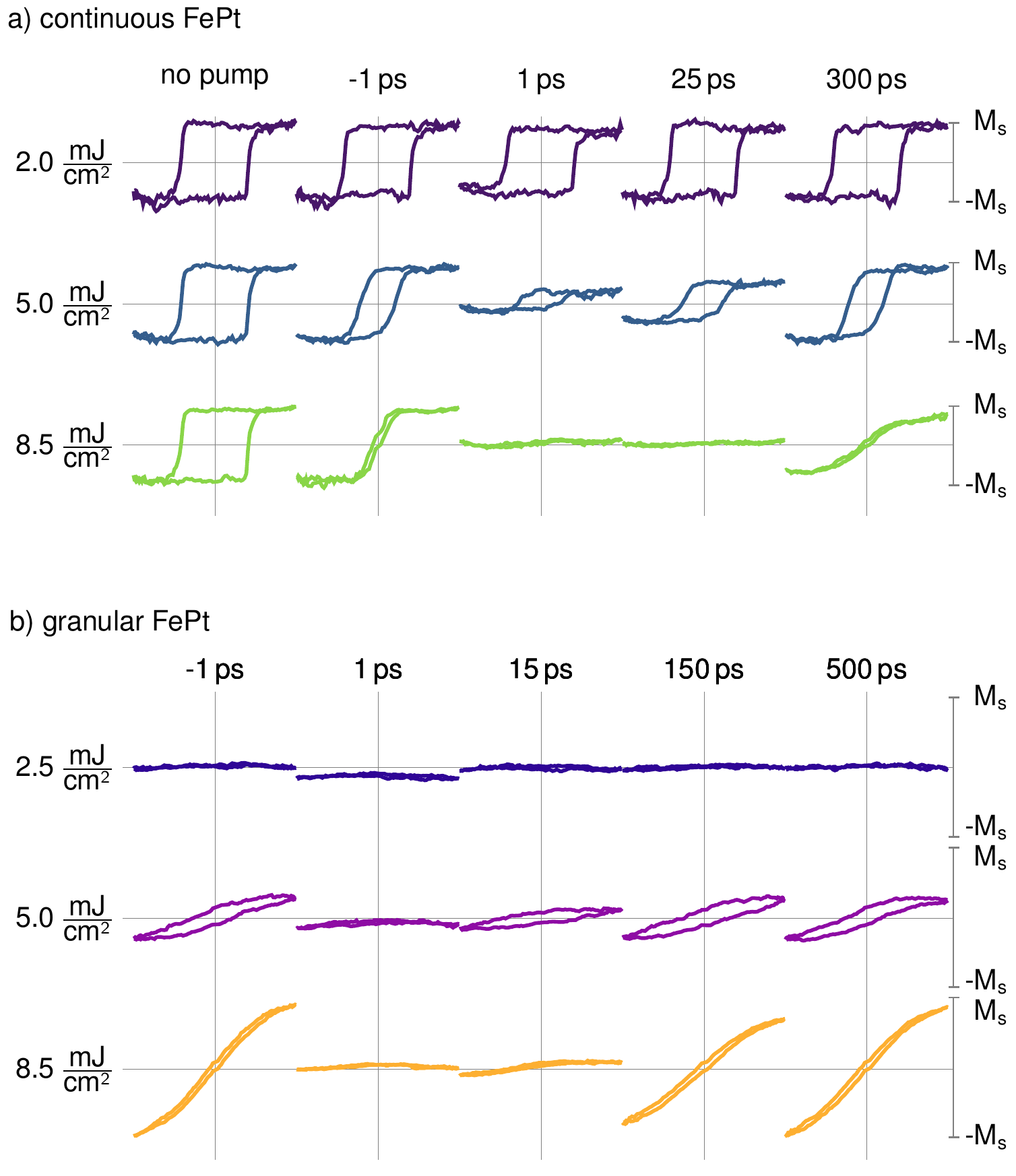}
	\caption{Hysteresis measured in the continuous (a) and granular (b) samples at different pump fluence and delay between the pump and the probe pulses. In all measurements the applied $B_\text{ext}$ ranges between $\pm0.75$\,T. The vertical scaling is fixed for all hysteresis curves and the saturation magnetization is normalized as described in the text a) to the hysteresis loop without pump and b) by the signal observed at the highest fluence.}
	\label{fig:hyst}
\end{figure}

In order to quantify the vertical axis of the TR-MOKE traces and to study the ultrafast magnetization dynamics in more detail, we performed hysteresis measurements on both samples using a maximum field of 0.75\,T at different pump fluences and delays between the pump and the probe pulses (Figure.~\ref{fig:hyst}). The two samples exhibit very different fluence and time-dependent hysteresis loops. Let us first focus on the continuous film for which the amplitude of the hysteresis was normalized to the one measured without excitation (Fig.~\ref{fig:hyst}(a)). The excitation at a low fluence of $F_\text{pump} = $2.0\,mJ/cm$^{2}$ shows that the absorbed energy of the pump pulse induces only a  very small decrease in the saturation magnetization at short timescales ($t=1$\,ps) without any change of the coercive field. At a medium fluence of $F_\text{pump}=5.0$\,mJ/cm$^{2}$, the saturation magnetization is reduced significantly, and the coercive field is smaller compared to the one measured without excitation. At high fluence of $F_\text{pump}=8.5$\,mJ/cm$^{2}$ the hysteresis is fully closed over a significant amount of time of at least 25\,ps.
For the granular sample, no hysteresis loop is observed at the low fluence of $F_\text{pump}=2.5$\,mJ/cm$^{2}$ (Fig.~\ref{fig:hyst}(b)), since the coercive field (cf. Fig.~\ref{fig:static}\,(a)) is larger than the maximum applied field of 0.75\,T. At this low fluence, the transient hysteresis at the time delay of 1\,ps shows the same reduction of the magneto-optical signal for the full $B_\text{ext}$ field range. This vertical shift indicates a reduced transient magnetization. No grains are switched in their magnetization when $B_\text{ext}$ is reversed and we only access the upper hysteresis branch of the granular film. A clear open hysteresis is observed at negative time delays for a medium fluence of $F_\text{pump} = 5.0$\,mJ/cm$^{2}$ and by further increasing $F_\text{pump}$ its amplitude becomes more and more pronounced and its coercivity is continuously reduced. The increasing hysteresis amplitude with increasing $F_\text{pump}$ implies  a larger fraction of switched particles. This phenomenon is related to the size distribution of the FePt grains, which leads to a large spread in the temperature changes proportional to the inhomogeneous light absorption that varies between 10 and 30$\%$ according to finite element simulations of the field enhancement effects in the optical absorption\,\cite{Granitzka2017}. This inhomogeneous distribution of heat cannot be washed out by electronic heat transport through the insulating carbon matrix, which rapidly cools down the FePt particles but does not transport the heat efficiently from grain to grain. Only the grains which experience a temperature rise close to the Curie temperature $T_\text{C}$ participate in the switching. We estimate that more than 90$\%$ of the particles are switched at 8.5 mJ/cm$^2$, since for 11 mJ/cm$^2$ the demagnetization only increases marginally. As a good approximation we thus calibrate the vertical axis in Fig.~\ref{fig:hyst}(b) by the amplitude of the hysteresis loop measured at negative delay with high fluence since this value is the saturation magnetization of the granular film. This calibration in Fig.~\ref{fig:hyst}(b) emphasizes that more and more particles switch with increasing $F_\text{pump}$. We mention that at the time delay of 1\,ps and for $F_\text{pump} = 8.5$\,mJ/cm$^2$ the hysteresis measured  in both -- continuous and granular -- films are flat and have zero amplitude, which indicates that for both films the temperature exceeds $T_\text{C}$ and therefore they are in the paramagnetic phase. On the other hand, we observe that for time delays larger than $\cong$15\,ps for the granular film and $\cong$25\,ps for the continuous one the hysteresis amplitudes gradually increase (Fig.~\ref{fig:hyst}), nicely visualizing the remagnetization dynamics of the films.

\begin{figure}
	\includegraphics[width = 8.6cm]{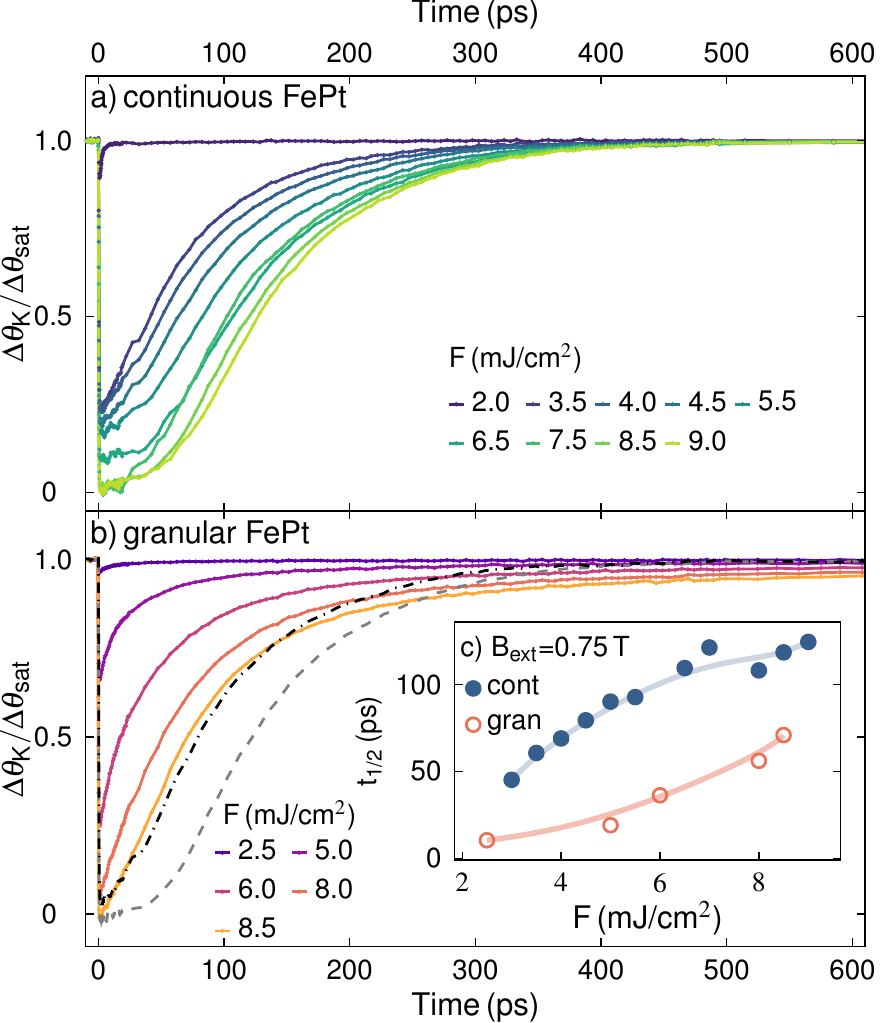}
	\caption{Pump fluence dependence of laser induced ultrafast magnetization dynamics. (a,b) $\Delta\theta_{K}$ measured in the (a) continuous and (b) granular samples as a function of the pump fluence at high external magnetic field of $B_\text{ext}=0.75$\,T. The dashed dotted and dashed lines in (b) are the $\Delta\theta_\text{K}$ signal induced in continuous film by $F_\text{pump}$ of 4.5 and 8.5\,mJ/cm$^2$, respectively. The inset (c) shows the pump fluence dependence of the time $t_\text{1/2}$ at which half of the demagnetization amplitude has recovered for 0.75\,T with the solid lines as guide to the eye.}
	\label{fig:fluence}
\end{figure}

To better illustrate the remagnetization dynamics as a function of the pump fluence, TR-MOKE measurements at selected $F_\text{pump}$ are shown for the continuous and the granular film in Fig.~\ref{fig:fluence}(a) and Fig.~\ref{fig:fluence}(b), respectively. The data were recorded for an applied field of $0.75$\,T. In both cases an increasing pump fluence causes the remagnetization to slow down. In addition, with the exception of the lowest fluence measurement of both samples, the granular film recovers significantly faster than the continuous film at equal incident fluence values. To illustrate this feature, Fig.~\ref{fig:fluence}(c) shows the pump fluence dependences of the time $t_\text{1/2}$ of the continuous and granular film. To directly visualize the faster recovery of the granular film, we plot by a dash-dotted line in Fig.~\ref{fig:fluence}(b) the $\Delta\theta_\text{K}$(t) measured for the continuous film at $F_\text{pump} = 4.5$\,mJ/cm$^{2}$. The early dynamics of the granular film are similar to this dash-dotted line when excited at almost twice the fluence. For 8.5\,mJ/cm$^{2}$ excitation of the continuous film the dynamics are much slower (dashed line), suggesting a significantly reduced light absorption in the granular film. However, for both fluences the continuous film approaches the saturation magnetization faster than the granular one beyond 300\,ps time delay (see Fig.~\ref{fig:fluence}(b)).
The similarity between the initial remagnetization of the continuous film and the granular one when exposed to twice the incident fluence suggests that a difference in the absorption for two films plays a role on the observed behavior. In order to examine the validity of this hypothesis, we have employed a transfer matrix calculation to estimate the optical absorption profile for the two samples\,\cite{Legu2013a}. We have used in our numerical calculation experimental values for the optical constants of the continuous (n= 3.30 + 2.63i) and granular (n= 2.98 + 1.78i) FePt films\,\cite{Cen2013,Granitzka2017}, which are measured in samples with comparable thickness, size of particles and carbon matrix. We note that the smaller extinction coefficient of the granular film (i.e. the imaginary part) not only indicates that it absorbs less energy compared to the continuous FePt layer. It also reveals that the absorption in the FePt grains is considerably larger than in the carbon, since otherwise the effective medium should have an increased imaginary part. %For an extensive estimation we include the 1\,nm Al layer which covers the continuous film in the numerical calculations.
The continuous FePt layer absorbs approximately $A_\text{cont}\cong 33\%$ of the incident fluence $F_\text{pump}$, while the granular one absorbs approximately $A_\text{gran}\cong 27\%$ in the effective medium that consists of FePt immersed in amorphous carbon. %To further evaluate the heat generated in the continuous FePt layer, we note that the electronic specific heat of FePt is about three times larger than the one for Al \textcolor{red}{\textbf{[cite?]}}. For that reason, we can approximate that the total energy of $A_\text{cont}=37$\% absorbed by the continuous sample initially heats up the FePt layer\,\cite{lin2008a, Scheid2019}.
Thus, the absorbed light energy heating the electron system of the granular FePt is only about $A_\text{gran}/A_\text{cont}=81$$\%$ of the continuous film.

Heat conduction to the surrounding carbon matrix is a second factor which reduces the energy density in the FePt grains. A four temperature model was recently proposed to capture this energy transfer for the nano-granular FePt-carbon composite\,\cite{Reid2018}. Due the comparable specific heat per volume of both materials, we estimate that the fraction of energy that will flow from FePt grains ($c_\text{V}=3.8\cdot 10^6$\,J/m$^3$K) to the carbon matrix is roughly proportional to its volume fraction of 0.3. Since the Debye temperature of amorphous carbon is considerably above room temperature\,\cite{Krumhansl1953}, $c_\text{V}$ increases from $c_\text{V}=2$ to 3.5$\cdot 10^6 $\,J/m$^3$K between room temperature and 800\,K and thus becomes comparable to the FePt value for our strong excitation regime. Hence, combining the two arguments, after heat flow to the carbon matrix the average FePt unit cell of the granular film contains only a fraction $\cong$70\% of the absorbed energy, which is, moreover only $\cong$81\% of the absorbed energy in the continuous film. This reconciles that about only half ($57\%$) of the incident fluence is needed in the continuous film to trigger the same magnetization dynamics as in the granular one.

Finally, we address the observation that for long timescales beyond 300\,ps the continuous film always approaches the saturation magnetization faster than the granular film, although it has absorbed twice the energy. The relatively weak van-der-Waals bonds of carbon was shown to exhibit a significantly reduced interface conductance as compared to metal-oxide interfaces with strong binding. \cite{Hsieh2011} Therefore, the carbon matrix should store the heat energy longer than the FePt and it can serve as an additional heat bath, which heats the FePt grains on long timescales up to 1\,ns and beyond. The temperature gradient across the interface has then reversed compared to the initial situation, where carbon cools the FePt particles, after they have been optically heated to very high temperatures.

\subsection{Conclusion}
We have compared the laser induced magnetization dynamics of a continuous and a granular FePt thin film with a similar thickness of about 10\,nm under various excitation fluences and external magnetic fields. Our experimental results show that the granular nature of the film influences the observed dynamics in several ways. First of all, the laser energy absorbed by the grains shows a broad distribution, where the average FePt unit cell in the continuous sample absorbs about twice the energy compared to the granular sample. Moreover, the carbon matrix changes the dynamics in three ways: It i) rapidly takes up approximately 30$\%$ of the absorbed energy and thus initially speeds up the remagnetization, ii) gives heat back to FePt after cooling and therefore slows down the remagnetization at later times iii) the grain boundaries prevent domain wall motion and therefore strongly reduce the impact of an external field on the remagnetization dynamics. We believe that this thorough comparison of the two morphologies of the L$1_0$ phase of FePt is useful as a reference for the laser-induced magnetization dynamics, especially on the timescale of remagnetization and cooling.

\subsection{Acknowledgements}
We acknowledge the BMBF for the financial support via 05K16IPA and the DFG via BA 2281/11-1. MD thanks the Alexander von Humboldt foundation for financial support.

%\bibliography{Literature1}
%

\end{document}